\documentclass[aps,prl,reprint,superscriptaddress,showpacs]{revtex4-1}
\usepackage{graphicx}
\usepackage{url}
\usepackage{color} 
\begin{document}

% Use the \preprint command to place your local institutional report
% number in the upper righthand corner of the title page in preprint mode.
% Multiple \preprint commands are allowed.
% Use the 'preprintnumbers' class option to override journal defaults
% to display numbers if necessary
%\preprint{}

%Title of paper
\title{Unwinding relaxation dynamics of polymers}

% repeat the \author .. \affiliation  etc. as needed
% \email, \thanks, \homepage, \altaffiliation all apply to the current
% author. Explanatory text should go in the []'s, actual e-mail
% address or url should go in the {}'s for \email and \homepage.
% Please use the appropriate macro foreach each type of information

% \affiliation command applies to all authors since the last
% \affiliation command. The \affiliation command should follow the
% other information
% \affiliation can be followed by \email, \homepage, \thanks as well.
\author{J.-C. Walter}
\affiliation{Instituut-Lorentz, Universiteit Leiden, P.O. Box 9506, 2300 RA Leiden, The Netherlands}
\affiliation{Institute for Theoretical Physics, KULeuven, Celestijnenlaan 200D, Leuven, Belgium}
\author{M. Baiesi}
\affiliation{Dipartimento di Fisica e Astronomia, Universit\`a di Padova, Via Marzolo 8, Padova, Italy}
\affiliation{INFN, Sezione di Padova, Via Marzolo 8, Padova, Italy}
\author{G. T. Barkema}
\affiliation{Institute for Theoretical Physics, Utrecht University, the Netherlands}
\affiliation{Instituut-Lorentz, Universiteit Leiden, P.O. Box 9506, 2300 RA Leiden, The Netherlands}
\author{E. Carlon}
\affiliation{Institute for Theoretical Physics, KULeuven, Celestijnenlaan 200D, Leuven, Belgium}
%\email[]{Your e-mail address}
%\homepage[]{Your web page}
%\thanks{}
%\altaffiliation{}

%Collaboration name if desired (requires use of superscriptaddress
%option in \documentclass). \noaffiliation is required (may also be
%used with the \author command).
%\collaboration can be followed by \email, \homepage, \thanks as well.
%\collaboration{}
%\noaffiliation

\date{\today}

\begin{abstract}
The relaxation dynamics of a polymer wound around a fixed obstacle
constitutes a fundamental instance of polymer with twist and torque and
it is of relevance also for DNA denaturation dynamics. We investigate
it by simulations and Langevin equation analysis. The latter predicts
a relaxation time scaling as a power of the polymer length times a
logarithmic correction related to the equilibrium fluctuations of the
winding angle.  The numerical data support this result and show that at
short times the winding angle decreases as a power-law.  This is also
in agreement with the Langevin equation provided a winding-dependent
friction is used, suggesting that such reduced description of the system
captures the basic features of the problem.
\end{abstract}

% insert suggested PACS numbers in braces on next line
% insert suggested keywords - APS authors don't need to do this
%\keywords{}

%\maketitle must follow title, authors, abstract, \pacs, and \keywords
\maketitle

The dynamics of polymers subject to spatial or topological constraints
has received quite some attention in recent years.  Interesting examples
are the translocation of DNA from a narrow pore (for a recent discussion
see e.g.~\cite{rowg11} and references therein) or the dynamics of
supercoiled DNA (see e.g.~\cite{crut07_sh}).  An important question is
whether the complex polymer dynamics can be described by a simple
equation of motion, using a one-dimensional reaction coordinate. This
issue arises, for instance, in the context of polymer translocation
(from a pore in a wall)
where it was shown that the Langevin equation fails to reproduce
simulation results~\cite{kant04}. This failure motivated extensive
studies. Various models were put forward, as the generalized Langevin
equation with a memory kernel~\cite{panj07}, or a deterministic two-phase
model~\cite{saka10}.  

The aim of this Letter is to study analytically and numerically
the unwinding relaxation dynamics, which also belongs to the above
class of problems. The equilibrium winding angles for polymers were
intensively studied in the past~\cite{rudn88,dupl88c,dros96}. These
studies are relevant for a series of problems in physics,
as e.g.~models for the behavior of flux lines in high-Tc
superconductors~\cite{dros96}. The relaxation dynamics of unwinding has
been much less studied~\cite{baum86,baie10a}, though it is a problem of
relevance in DNA melting dynamics, but also as a fundamental issue of
polymer dynamics involving twist and torsion.

\begin{figure}[!t]
\includegraphics[angle=0,width=8cm]{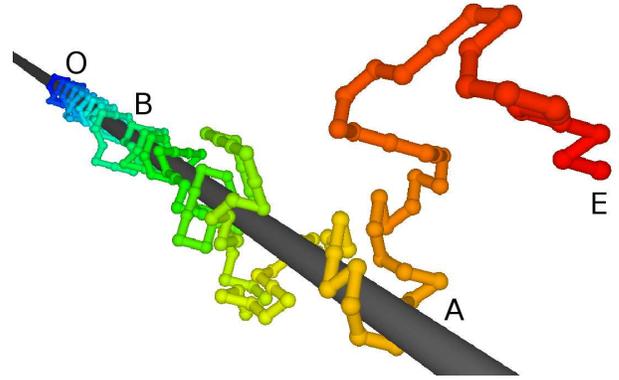}
\caption{(Color online) Snapshot of a polymer configuration on the fcc lattice,
during the unwinding from a bar. The winding angle is defined as $\theta =
\sum_{i=1}^L \Delta \theta_{i,i+1}$, where $\Delta \theta_{i,i+1}$
is the difference in angles between monomer $i+1$ and $i$ measured
with respect to the bar.  The hue follows the monomers order from $i=1$
(``O'', attached to the bar, blue online) to $i=L$ (``E'', red online).
The configuration displays a tightly bound helix (OB), a loose helix
(BA), and a free end (AE).
}
\label{fig:1}
\end{figure}

We consider a polymer initially wound around a long impenetrable bar
(see Fig.~\ref{fig:1}), to which it is attached at one end. Since this is
an entropically highly unfavorable situation, the polymer will unwind,
starting at the loose end; and given enough time, it will relax towards
the equilibrium state in which it is no longer winding around the bar. To
monitor the unwinding process, we keep track of the winding angle $\theta$
of the last monomer of the polymer, which measures the angle accumulated
by the chain around the bar from the first attached monomer to the last
free one.  We treat the case of polymers with internal excluded volume
by studying a self-avoiding walk (SAW) and we support our arguments by
also investigating the motion of a random walk (RW). Compared to the
more complex unwinding of a double stranded DNA helix, the advantage
of dealing with a single polymer around a fixed obstacle is that this
winding angle provides a well-defined ``reaction coordinate''.

%%%%%%%%%%%%%%%%%%%%%%%  FIG01  %%%%%%%%%%%%%%%%%%%%%%%%%%%%%%%%%%%%%%%%%%%%%
%%%%%%%%%%%%%%%%%%%%%%%  FIG01  %%%%%%%%%%%%%%%%%%%%%%%%%%%%%%%%%%%%%%%%%%%%%

\begin{figure}[!t]
\includegraphics[angle=0,width=8cm]{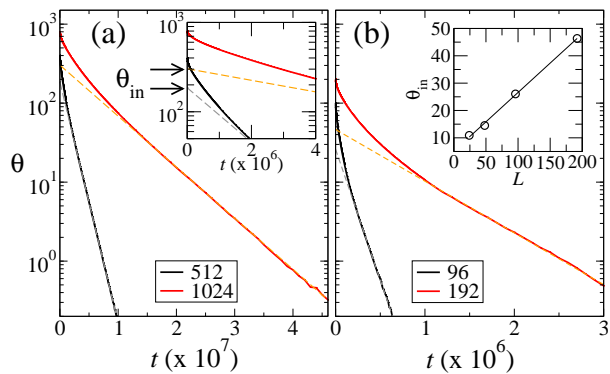}
\caption{(Color online) Simulations (solid lines)
of average winding angles vs. time in a semi-logarithmic plot for (a)
RW in two dimensions and (b) SAW in three dimensions. 
Only data for two polymer lengths are shown (see legends). 
Dashed lines are
fit of the exponential decay at long times. Inset of (a): zoom-in of 
the short-time behavior with extrapolated intercepts $\theta_{in}$.
Inset of (b): plot of $\theta_{in}$ vs. polymer length $L$, for the
SAW data, showing a linear behavior $\theta_{in} \sim L$.}
\label{fig:3}
\end{figure}

The numerical calculations were performed using lattice polymers,
specifically $L$-step RWs on a square lattice and SAWs on a
face-centered-cubic (fcc) lattice. An update consists of a local corner
flip or an end-flip move (Rouse dynamics), and a time step includes
$L$ updates at random locations.  The initial configuration for a RW is
constructed by the repetition of a sub-walk of $8$ monomers winding around
the $(0,0)$ site (representing the bar), with a resulting initial winding
angle $\theta_0 = \pi L/4$.  Similarly, for the SAW on the fcc lattice
we repeat a helix formed by $6$ steps ($\theta_0 = \pi L/3$) around
the bar in the direction $(1,1,0)$.  Figure~\ref{fig:3} shows a plot of
$\theta$ vs. time in a semi-logarithmic scale obtained from numerical
simulations: one distinguishes a long-time regime where
$\theta$ relaxes exponentially and a short-time regime that deviates
from the exponential decay. We will discuss the two cases separately.

%%%%%%%%%%%%%%%%%%  FIG03 %%%%%%%%%%%%%%%%%%%%%%%%%%%%%%%%%%%%%%%%%%%%%%
%%%%%%%%%%%%%%%%%%  FIG03 %%%%%%%%%%%%%%%%%%%%%%%%%%%%%%%%%%%%%%%%%%%%%%

Our analytical scheme is based on a one-dimensional Langevin
equation for the variable $\theta$:
\begin{equation}
\gamma_\tau \frac{d\theta}{dt} = 
- \frac{\partial {\cal F}(\theta, L)}{\partial \theta} + \eta\,,
\label{langevin}
\end{equation}
with $\cal F$ the equilibrium free energy for a polymer of length $L$
and winding angle $\theta$, $\gamma_\tau$ the torque friction and $\eta$
a noise term.  The main focus is on the time evolution of the average
winding angle $\langle \theta \rangle$ (indicated with $\theta$ for
simplicity) and not in fluctuations, so the noise term will be neglected.

The degrees of freedom parallel to the bar are not relevant for a RW
and one can restrict the study to a two-dimensional walk, with the bar
replaced by an excluded site. For a planar RW the free energy is known
exactly~\cite{rudn88}:
\begin{equation}
{\cal F}_{RW}(\theta, L) = - k_B T \log \left[{\rm cosh}^{-2} 
\left( \frac{\pi \theta}{\log L}\right) \right]\,.
\label{2dfree}
\end{equation}
For a SAW wound around the bar recent numerical
simulations suggest a similar scaling form~\cite{walt11b}:
\begin{equation}
{\cal F}_{SAW}(\theta, L) = - k_B T \log 
\left[ p \left( \frac{\theta}{(\log L)^{0.75}  }\right)\right]\,,
\label{3dfree}
\end{equation}
where $p()$ is the probability distribution of winding angles obtained
from equilibrium Monte Carlo sampling. Here the exponent $0.75$ is a
numerical estimate~\cite{walt11b}. Since both free
energies involve a scaling variable $\theta/(\log L)^\alpha$,
with $\alpha=1$ for RWs and $\alpha \approx 0.75$ for SAWs, we can analyze
the two processes on equal footing.

We focus first on the longest relaxation time. Eqs.~(\ref{2dfree}) and
(\ref{3dfree}) are quadratic for small $\theta$. Hence using the
lowest-order term and neglecting other proportionality factors, for small angles
one obtains an equation of the form
\begin{equation}
\frac{d\theta}{dt} \propto \frac{-\theta}{\gamma_\tau (\log L)^{2\alpha}}\,.
\label{diff2}
\end{equation}
In order to gain some insights on the $L$-dependence of the torque
friction $\gamma_\tau$ one can consider a particle rotating at a fixed
distance $R$ from an origin and subject to a constant tangential force
$f$. The Langevin equation in $\theta$ is of the form $\gamma_\tau
\frac{d\theta}{dt} = \Omega = f R$, where $\Omega$ is the torque. The
equation can be transformed into a cartesian coordinate $x=R\theta$,
yielding
\begin{equation}
f = \frac{\gamma_\tau}{R^2} \frac{Rd \theta}{dt} =
\frac{\gamma_\tau}{R^2} \frac{dx}{dt} = 
\gamma \frac{dx}{dt}\,,
\label{diff_eq_lin}
\end{equation}
where $\gamma$ is the friction associated with linear displacement. 
This implies that $\gamma_\tau = R^2 \gamma$.  
By integrating over $L$ monomers we obtain $\gamma\sim L$ and hence 
an average torque friction $\gamma_\tau \sim L^{1+2\nu}$, where
the Flory exponent $\nu$ describes the average end-to-end squared
distance $\langle R^2 \rangle \sim L^{2\nu}$ for a polymer in equilibrium:
$\nu=1/2$ for a RW, while $\nu \simeq 0.588$ for a three-dimensional SAW.
Plugging the estimated $\gamma_\tau$ into Eq.~(\ref{diff2}) one finds
the following relaxation time-scale
\begin{equation}
\tau_L \sim L^{1+2\nu} (\log L)^{2\alpha}\,.
\label{tauL}
\end{equation}
If hydrodynamic effects are included, the friction for linear displacement
grows as $\gamma\sim L^\nu$, and the relaxation time becomes $\tau_L
\sim L^{3\nu} (\log L)^{2\alpha}$.  Note that the leading term of
Eq.~(\ref{tauL}) is similar to the Rouse time $\tau_L^{\rm Rouse}
\sim L^{1+2\nu}$, which is the equilibration time of a free polymer
\cite{doi89}. This is also a lower bound for the unwinding relaxation
time, i.e.~$\tau_L \geq \tau_L^{\rm Rouse}$, as the attachment to the
bar and its steric hindrance are unlikely to speed up the equilibration
process.

In the simulations we determined the total unwinding time $\tau^*_L$,
i.e. the average time needed for the unwinding process to be
completed. We defined it as the time it takes to reach $\theta=0$ for the first time. As the
polymers are initially wound to $\theta_0 \sim L$, one has to take into
account that the relaxation starts from a higher winding angle for longer
polymers. The analysis of the numerical data (see Fig.~\ref{fig:3}) shows
that the asymptotic decay is well-fitted by $\theta (t)= \theta_{in}
\exp(-t/\tau_L)$ and the intercept $\theta_{in}$ scales linearly with $L$.
Hence the condition $\theta \left( \tau^*_L \right) \sim 1$ gives
\begin{equation}
\tau^*_L \sim \tau_L \log L \sim L^{1+2\nu} (\log L)^{2\alpha+1}\,,
\label{taustar}
\end{equation}
Thus, the total unwinding time $\tau^*_L$ differs by a factor $\log L$ from 
the relaxation time-scale $\tau_L$.

%%%%%%%%%%%%%%%%%%  FIG02  %%%%%%%%%%%%%%%%%%%%%%%%%%%%%%%%%%%%%%%%%%%%%%
\begin{figure}[!t]
\includegraphics[angle=0,width=8cm]{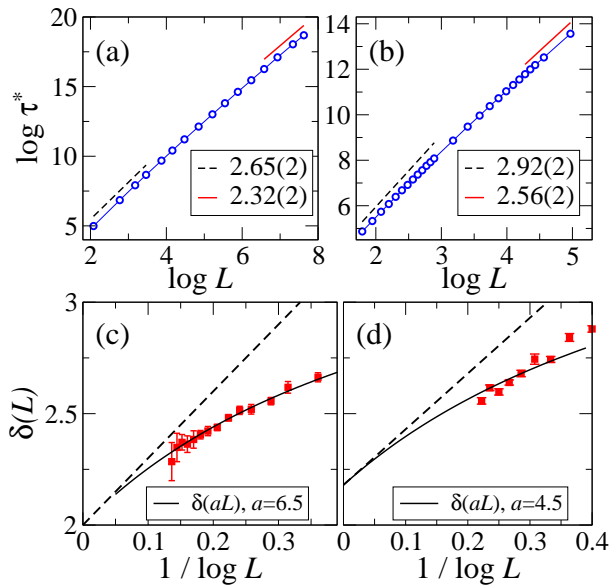}
\caption{(Color online) (a) Plot of $\log \tau^*_L$ vs. $\log L$ for the
RW, with lengths up to $L=2048$.  Averages are from $10^4$ independent
runs for $L<200$ down to $200$ for $L=1536,\,2048$. Dashed and solid
lines are fits in the short and long $L$ regimes, showing a systematic
variation in the exponent.  (b) Same as (a) for SAWs, with lengths up
to $L=144$ ($10^5$ independent runs per $L$).  (c)
Squares: Plot of the RW running exponent $\delta(L)$ obtained from
simulations [estimated by a centered difference of data in (a)]
as a function of $1/\log L$.  Dashed and solid lines represent the
scaling $\delta(L)$ and $\delta(a_{RW} L)$, with $a_{RW} = 6.5$, from
Eq.~(\ref{eff_exp_LE}), respectively. (d) The same for SAWs data from
panel (b), with $a_{SAW}=4.5$.}
\label{fig:2}
\end{figure}
%%%%%%%%%%%%%%%%%%  FIG02  %%%%%%%%%%%%%%%%%%%%%%%%%%%%%%%%%%%%%%%%%%%%%%

Plots of $\log \tau^*_L$ vs. $\log L$ are shown in Fig.~\ref{fig:2}(a)
(RW) and \ref{fig:2}(b) (SAW).  In order to analyze the data appropriately
we computed $\delta(L)$, defined as the ``local" slope in the $\log \tau^*_L$
vs. $\log L$ plot for a given size $L$. Eq.~(\ref{taustar}) implies
\begin{equation}
\delta(L) \equiv \frac{d \log \tau^*_L}{d \log L} = 
1+2\nu+ \frac{2\alpha+1}{\log L}\,,
\label{eff_exp_LE}
\end{equation}
Figures~\ref{fig:2}(c) and (d) (squares) show the numerical estimates
of $\delta(L)$ vs. $1/\log L$ for the RW and SAW, respectively. The
asymptotic scaling predicted by Eq.~(\ref{eff_exp_LE}) implies a
straight line for $\delta(L)$ when plotted as a function of $1/\log L$
(dashed lines in Figs.~\ref{fig:2}(c) and (d)). The curvature in the data
indicates that further finite-size corrections should be included. In
order to rationalize them we introduce a finite-size scaling ansatz
$\delta(aL) = 1+2\nu+ (2\alpha+1)/\log(aL)$ in which an amplitude ``$a$"
is included in the logarithmic factor as a single
fitting parameter. The best fit of $\delta(aL)$ to the data points
produces the two solid lines in Figs.~\ref{fig:2}(c) and (d). The data
for the RW (c) are in excellent agreement with the ansatz, while the
SAW data are less conclusive: they involve much heavier computations
and are thus restricted to much shorter polymers.

%%%%%%%%%%%%%%%%%%  FIG04  %%%%%%%%%%%%%%%%%%%%%%%%%%%%%%%%%%%%%%%%%%%%%%
\begin{figure}[t]
\includegraphics[angle=0,width=7.8cm]{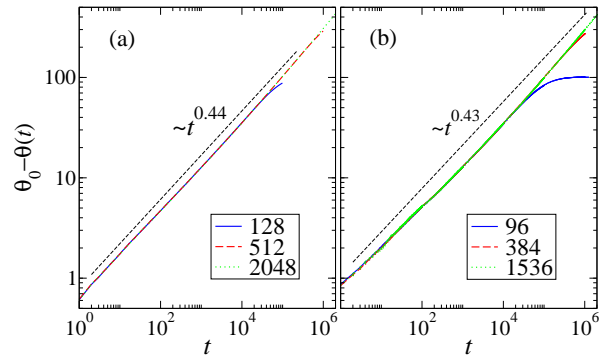}
\caption{(Color online) $\theta_0 - \theta(t)$ versus
$t$ for the RW (a) and  the SAW (b). The log-log
scale highlights the short-time regime, which behaves as a power law
(Eq.~(\ref{powerlaw_rho})) with $\rho \simeq 0.44$ (RW) and $\rho\simeq 0.43$
(SAW).}
\label{fig:4} 
\end{figure}
%%%%%%%%%%%%%%%%%%  FIG04  %%%%%%%%%%%%%%%%%%%%%%%%%%%%%%%%%%%%%%%%%%%%%%

We consider next the early-time dynamics of $\theta$.  Figure~\ref{fig:4}
shows a plot of $\theta_0 - \theta(t)$ vs.~$t$ in log-log scale.
For $t \lesssim 10^4$ the data are fitted by a power-law behavior
\begin{equation}
\theta_0 - \theta(t) \sim t^\rho\,,
\label{powerlaw_rho}
\end{equation}
with $\rho\approx 0.44$ and $\rho\approx 0.43$ for RW and SAW,
respectively. To understand this behavior we consider again the Langevin
equation (\ref{langevin}). At very high winding, where $\theta(t) \sim
\theta_0 \sim L$ the torque due to free energies in Eqs.~(\ref{2dfree})
and (\ref{3dfree}) are of little use as they describe equilibrium
fluctuations for small $\theta$'s.  In the early stages of the dynamics
we expect unwinding only near the free end, regardless of the polymer
length.  The decrease of $\theta$ is then linearly related to the
length of the unwound part of the polymer, and to leading order also
linearly to the increase in entropy.  We thus assume that the torque is 
constant ($L$-independent), $\tau_0 =- \frac{\partial
{\cal F}}{\partial \theta} = $ {\it const}.  At high winding the friction decreases,
as the part of the polymer which is tightly wound around the bar does not
contribute to it. For $\theta_0 - \theta \ll L$, the friction coefficient
should depend only on the difference $\theta_0 - \theta$.  Let us consider
a friction coefficient vanishing as a power-law as $\gamma_\tau (\theta)
\sim \left( \theta_0 - \theta \right)^x$. Combining this ansatz for
$\gamma_\tau$ with the argument for a constant torque $\tau_0$, from
(\ref{langevin}) one obtains
\begin{equation}
-\gamma_\tau (\theta) \frac{d\theta}{dt} \sim (\theta_0 - \theta)^x 
\frac{d}{dt} \left(\theta_0 - \theta\right) \sim \tau_0\,,
\end{equation}
which integrated in time, and using the initial condition $\theta
(0) = \theta_0$, yields a power-law scaling as that given in
Eq.~(\ref{powerlaw_rho}) with $\rho={1/(1+x)}$.

\begin{figure}[!tb]
\includegraphics[angle=0,width=7.5cm]{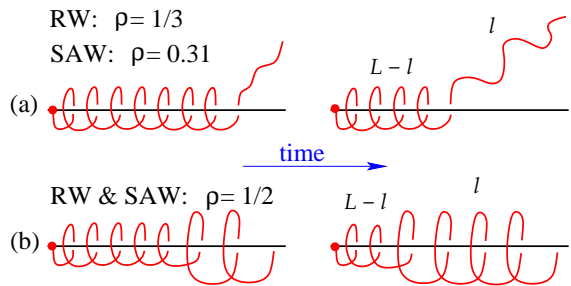}
\caption{Two possible configurations of polymers during unwinding.  (a)
A tight helix of length $L-l$ connected to a loose equilibrated end of
length $l$. (b) The part of the polymer detached from the bar here has
still some winding. The exponents $\rho$ governing the early-time decay
of the winding angle predicted in the two cases are given.}
\label{fig:5} 
\end{figure}

To estimate the exponent $x$ we introduce two different types of
hypotheses about the shape of the polymer in the early stages of
unwinding. These are sketched in Fig.~\ref{fig:5}(a) and (b). In the case
(a), we consider a tightly wound polymer for a length $L-l$ and an unwound
loose part of length $l$ and assume that the latter is equilibrated.
We denote the winding per unit length in the wound part with $\Delta
\omega_1$ and that of the loose part with $\Delta \omega_2$ ($\Delta
\omega_2\approx 0$ in the case of Fig.~\ref{fig:5}(a)).  The winding angle
$\theta = (L-l) \Delta \omega_1 = \theta_0 - l \Delta \omega_1$, hence
$\theta_0 - \theta = l \Delta \omega_1$. As shown above in the discussion
of the late-time relaxation, an equilibrated polymer of length $l$ has a
torque friction scaling as $l^{1+2\nu}$, therefore $\gamma_\tau (\theta)
\sim (\theta_0 - \theta)^{1+2\nu}$ which implies $\rho = 1/(2\nu+2) =
1/3$ for a RW and $\rho = 0.31$ for a SAW. An alternative conformation
is shown in Fig.~\ref{fig:5}(b).  In this case we consider a ``looser''
helix of length $l$ with density of winding per unit length $\Delta
\omega_2>0$ connected to a tightly wound helix of length $L-l$. Only
the former contributes to the friction. In addition we assume that the
looser helix does not change its radius and pitch in time (thus $\Delta
\omega_2$ is constant). This seems reasonable at least for the early
times of the dynamics.  We then have $\theta = (L-l) \Delta \omega_1 +
l \Delta \omega_2 = \theta_0 -l (\Delta \omega_1 - \Delta \omega_2)$. As
the loose helix maintains its shape while growing the friction is simply
proportional to its length: $\gamma_\tau (\theta) \sim l \sim (\theta_0 -
\theta)$, which yields $\rho=1/2$ both for a RW and a SAW.

%%%%%%%%%%%%%%%%%%  FIG05  %%%%%%%%%%%%%%%%%%%%%%%%%%%%%%%%%%%%%%%%%%%%%%
%%%%%%%%%%%%%%%%%%  FIG05  %%%%%%%%%%%%%%%%%%%%%%%%%%%%%%%%%%%%%%%%%%%%%%

The conformations of Fig.~\ref{fig:5} are of course ``idealized''
and should represent two extreme cases. In (a) the
loose end stretches out from the bar causing a more rapid increase in
the friction compared to (b).  The exponent $\rho=1/2$ predicted for
the case (b) is quite close to $\rho \approx 0.45$ found in simulations. 
Snapshots such as that in Fig.~\ref{fig:1}
suggest that the actual polymer conformations are hybrids of those in
Fig.~\ref{fig:5}. Starting from the free end, one notices a very
loose part which does not add much to the winding angle (segment AE in
Fig~\ref{fig:1}). This is reminiscent of the loose equilibrated end of
Fig.~\ref{fig:5}(a). There is then an intermediate part (BA 
in Fig.~\ref{fig:1}) wound around the bar, but not tightly,
resembling the loose helix of Fig.~\ref{fig:5}(b).

In conclusion, in this Letter we investigated numerically the relaxation
dynamics of polymers wound around a fixed obstacle and we have provided
an analytical scheme based on a Langevin equation for the winding angle.
Studying such equation in the late relaxation stage we predict the scaling
form of the friction and consequently of the unwinding time-scale, which
involves logarithmic corrections to the power-law of the chain length.
The same equation is also useful in the regime at short times, where a
friction depending on the unwinding is needed to describe the observed
scaling of the winding angle.  The two cases analyzed numerically, a SAW
and a RW, provide a consistent picture of the dynamical behavior. Although
logarithmic factors are notoriously difficult to study in simulations,
finite-size scaling extrapolations of our results are compatible with
the predictions of the Langevin equation. It is possible that such
strong corrections affect also the unwinding of two polymers from a
double-helical conformation. A recent numerical study~\cite{baie10a}
yields an unwinding time scaling as $\tau^*_L \sim L^{2.58}$; numerically,
this scaling is consistent with that of the running exponent found in this
work for the longest polymers (see Fig.~\ref{fig:2}(b)). It is thus
plausible that the relaxation time of an unwinding double-helix is
also captured by Eq.~(\ref{taustar}).  Besides delving new fundamental
aspects of polymer dynamics and providing a reference case for DNA
denaturation dynamics, this study may also serve as a basis for other
types of investigations involving rotational dynamics, as for instance
the relaxation of plectonemic structures which form in overtwisted
DNA. Modeling the statics and dynamics of DNA plectonemes has been of
recent great interest~\cite{crut07_sh,dani11,wada09,neuk11}.

We thank Helmut Schiessel for interesting discussions. 
This work is a part of the research program of the
 ``Stichting voor Fundamenteel Onderzoek der Materie'' (FOM), which
 is financially supported by the ``Nederland Organisatie
 voor Wetenschappelijk Onderzoek'' (NWO).

%  \bibliography{/home/staff/enrico/TEX/biblio}
%  \bibliography{/home/enrico/TEX/biblio}

\vspace{-5mm}

\end{document}